\documentstyle[12pt]{article}
\let\ssection=\section
\renewcommand{\section}{\setcounter{equation}{0}\ssection}

\newcommand\mathC{\mkern1mu\raise2.2pt\hbox{$\scriptscriptstyle|$}
        {\mkern-7mu\rm C}}        
\newcommand{\mathR}{{\rm I\! R}}  
\newcommand{\mathZ}{{\rm Z\!\! Z}}
\newcommand\Q{{\cal Q}}

\newcommand\K{{\cal K}}
\newcommand\id{{\rm id}}
\newcommand\ie{{\em i.e.}}

\newcommand\braket[2]{\langle#1|#2\rangle}
\newcommand\ket[1]{\,|#1\rangle}
\newcommand\Ob[1]{{\rm Ob}(#1)}
\newcommand\Diff[1]{{\rm Diff}(#1)}
\newcommand\Hom[1]{{\rm Hom}(#1)}

\newcommand\Ran[1]{{\rm Ran}\;#1}
\newcommand\Dom[1]{{\rm Dom}\;#1}
\newcommand\AF[1]{{\rm AF}(#1)}
\newcommand\eq[1]{Eq.\ (\ref{#1})}
\newcommand\eqs[2]{Eqs.\ (\ref{#1}--\ref{#2})}

\newcommand\mapdown[1]{\Big\downarrow
        \rlap{$\vcenter{\hbox{$\scriptstyle#1$}}$}}
\newcommand\mapright[1]{\smash{
        \mathop{\mbox{\large{$\;\longrightarrow\;$}}}\limits^{#1}}}
\newcommand\smapright[1]{\smash{
        \mathop{\mbox{$\;\longrightarrow\;$}}\limits^{#1}}}
\newcommand\mapleft[1]{\smash{
        \mathop{\mbox{\large{$\;\longleftarrow\;$}}}\limits^{#1}}}
\newcommand\bundle[3]{\begin{array}[t]{c}
        {#1}\\ \mapdown{#2}\\ {#3}\end{array}}

\newcommand\bundlemapLR[2]{\begin{array}[t]{c}\mapleft{#1}\\
            \phantom{\mapdown{}}\\\mapright{#2}\\\end{array}}

\begin{document}
\begin{titlepage}
\hspace{09truecm}Imperial/TP/03-04/??


\begin{center}
{\large\bf Quantising on a Category}\footnote{Dedicated with
respect to the memory of Jim Cushing}
\end{center}

\vspace{0.8 truecm}

\begin{center}
            C.J.~Isham\footnote{email: c.isham@imperial.ac.uk}\\[10pt]
            The Blackett Laboratory\\
            Imperial College of Science, Technology \& Medicine\\
            South Kensington\\
            London SW7 2BZ\\
\end{center}

\begin{abstract}
We review the problem of finding a general framework within which
one can construct quantum theories of non-standard models for
space, or space-time. The starting point is the observation that
entities of this type can typically be regarded as objects in a
category whose arrows are structure-preserving maps. This
motivates investigating the general problem of quantising a system
whose `configuration space' (or history-theory analogue) is the
set of objects $\Ob\Q$ in a category $\Q$.

We develop a scheme based on constructing an analogue of the group
that is used in the canonical quantisation of a system whose
configuration space is a manifold $Q\simeq G/H$, where $G$ and $H$
are Lie groups. In particular,  we choose as the analogue of $G$
the monoid of `arrow fields' on $\Q$. Physically, this means that
an arrow between two objects in the category is viewed as an
analogue of momentum. After finding the `category quantisation
monoid', we show how suitable representations can be constructed
using a bundle (or, more precisely, presheaf) of Hilbert spaces
over $\Ob\Q$. For the example of a category of finite sets, we
construct an explicit representation structure of this type.

\end{abstract}
\end{titlepage}

\section{Introduction}
\subsection{Motivational Remarks}
Attempts to construct a quantum theory of gravity provide many
challenges for quantum theory itself. Some of these involve
conceptual issues of the type in which Jim Cushing would certainly
have been interested. For example: is it meaningful to talk about
a quantum theory in the absence of any background spatio-temporal
structure? Or what is meant by a `quantum theory of the universe'
in quantum cosmology?

Other issues are of a more mathematical nature and concern the
technical challenge of finding a structure that can be understood
as a `quantum theory of space-time', or a `quantum theory of
space'. This is the focus of the present paper, which reviews
(hopefully, in a fairly pedagogical way) some recent work that was
stimulated by a desire to quantise discrete models of space or
space-time \cite{IshQCT1_03} \cite{IshQCT2_03} \cite{IshQCT3_03}.
An example of a discrete space-time is a `causal set' in which the
fundamental ingredient is the causal relations between space-time
points \cite{Sor91}. In mathematical terms, this means modeling
space-time with a partially-ordered set (`poset') $C$ where, if
$p,q\in C$ are such that $p\prec q$, then the space-time point $q$
lies in the  causal future of the space-time point $p$.

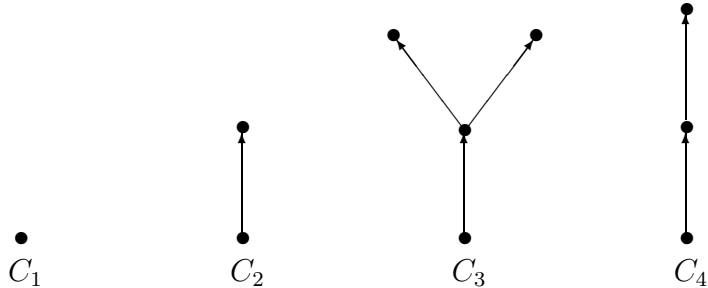
\begin{figure}[h]
\begin{picture}(100,100)(0,0)
\put(60,12){
\begin{picture}(80,100)(0,0)
    \put(0,0){$\bullet$}\put(-2,-13){$C_1$}
\end{picture}
\begin{picture}(80,100)(0,0)
\put(0,0){$\bullet$} \put(2.5,3){\vector(0,1){40}}
\put(0,42){$\bullet$}\put(-2,-13){$C_2$}
\end{picture}
\begin{picture}(80,100)(0,0)
\put(0,0){$\bullet$} \put(2.5,3){\vector(0,1){40}}
\put(0,41){$\bullet$} \put(2,45){\vector(-3,4){25}}
\put(4,45){\vector(3,4){25}}\put(-27,77){$\bullet$}
\put(27,77){$\bullet$}\put(-2,-13){$C_3$}
\end{picture}
\begin{picture}(100,100)(0,0)
\put(0,0){$\bullet$} \put(2.5,3){\vector(0,1){40}}
\put(0,42){$\bullet$} \put(2.5,48){\vector(0,1){40}}
\put(0,87){$\bullet$}\put(-2,-13){$C_4$}
\end{picture}
}
\end{picture}
\caption[]{A collection of four causal
sets}\label{Fig:Cat4objects}
\end{figure}

To illustrate some of the main issues, consider the `toy' model in
Figure \ref{Fig:Cat4objects}, which contains just four causal
sets. Can we find a mathematical framework for constructing
theories in which the universe could be in a quantum superposition
of these four basic space-times? For example, are there quantum
states $\ket{C_i}$, $i=1,2,3,4$, such that the general state can
be written as
\begin{equation}
    \ket\psi=\sum_{i=1}^4\alpha_i\ket{C_i}\label{psi=sum4CS}
\end{equation}
for complex coefficients $\alpha_i$, $i=1,2,3,4$?

Note that, if there is such a framework, then it must describe a
{\em history\/} theory. This is because a causal set represents a
{\em space-time\/}, and therefore could not be associated with a
quantum state in the standard approaches to quantum theory in
which a state is defined at a fixed moment in time. The most fully
developed theory of this type is the consistent histories approach
to quantum theory \cite{Gri84} \cite{Omn88} \cite{GH90}.

Thus one motivation for the work in the present paper is to
construct a framework for discussing quantum theories for a system
whose `history configuration space' is a specific collection of
causal sets. A similar problem in non-history quantum theory would
be to find a quantum framework for a system in which physical
space at each moment in time is represented by one of a collection
of topological spaces\footnote{From a purely mathematical
perspective these two examples are related. Indeed, any $T_0$
topological space gives rise to a partially ordered set in which
$x\leq y$ is defined to mean that $y\in\overline{\{x\}}$, the
closure of the set $\{x\}$. Conversely, any poset gives rise to a
topological space that is generated by the lower sets in the poset
\cite{Sor91b}. For example, moving from left to right in Figure
\ref{Fig:Cat4objects}, the posets describe a topology on a set
with 1,2,4 and 3 points respectively. Thus a theory of quantising
posets could be construed as a `canonical' theory of quantum
topology as well as a history theory of causal sets.}.

There has been much work on quantising a system whose classical
(canonical) configuration space is a differentiable manifold $Q$.
One particularly powerful approach is to associate with $Q$ an
algebra whose irreducible representations constitute quantisations
of the system. For example, if $Q$ is the real line $\mathR$, the
appropriate algebra is generated by the canonical commutation
relation
\begin{equation}
            [\,\hat x,\hat p\,]=i\hat 1.       \label{1DCCR}
\end{equation}

This raises the question of whether there is an analogue of such
an algebra for a quantum theory of discrete spaces or space-times:
for example, a space-time history theory based on a collection of
causal sets. As we shall see, the answer is `yes', providing one
adopts the approach to consistent history theories developed by
the author and collaborators in which propositions about the
history of the system are represented by projection operators on a
`history quantum state space' \cite{Isham94} \cite{IL94}.

In fact, the quantisation scheme we shall develop applies to far
more than just collections of causal sets, or topological spaces.
This new scheme involves constructing a quantisation algebra for a
system whose `configuration space' (or history-theory analogue
thereof) is the collection of objects in a {\em category\/}. The
arrows in the category are then interpreted as the analogue of
momentum variables.  For example, Figure \ref{Fig:Cat4objects}
illustrates a category with four objects, where the arrows between
a pair of causal sets are defined as order-preserving maps.

The invocation of category theory may sound rather abstract, but
the idea works well in practice and provides a set of powerful
tools for quantising types of physical system that are outside the
scope of existing methods.

The plan of the paper is as follows. We start by reviewing the
quantisation algebra for a system whose configuration space (or
history-theory equivalent) is a manifold. Then, in Section
\ref{SubSec:QuCSTops}, we motivate the idea of quantising a system
whose configuration space is the collection of objects in a
category. A key ingredient is the concept of an {\em arrow
field\/}, which is introduced in Section \ref{Sec:ArrowFields}.
Arrow fields play a central role in constructing the quantisation
algebra for a category, and this is explained in Section
\ref{Sec:ArrowFieldQT}, as are the techniques for constructing
representations of this algebra. Then, in Section
\ref{Sec:CategoryOfSets}, the general scheme is applied to the
special case of a category of sets with structure: for example,
causal sets, or topological spaces.

\section{Quantisation Algebras}
\label{Sec:QuAlgebras}
\subsection{A Non-Relativistic Particle}
\label{SubSec:QAR3} Let us consider first the quantum framework
for a non-relativistic point particle moving in three spatial
dimensions. The canonical commutation relations that specify the
quantisation are
\begin{eqnarray}
    [\,\hat x_i,\hat x_j\,]&=&0            \label{CCRxx}\\
    {[\,}\hat p_i,\hat p_j\,]&=&0          \label{CCRpp}\\
    {[\,}\hat x_i,\hat p_j\,]&=&i\delta_{ij}\hat 1    \label{CCRxp}
\end{eqnarray}
for $i,j=1,2,3$. On defining $\hat U({\bf a}):=e^{i{\bf
a}\cdot{\bf \hat p}}$, and $\hat V({\bf b}):= e^{i{\bf b}\cdot{\bf
\hat x}}$ for three-vectors ${\bf a},{\bf b}$, \eqs{CCRxx}{CCRxp}
give the exponentiated relations
\begin{eqnarray}
    \hat U({\bf a_1})\hat U({\bf a_2})&=&
        \hat U({\bf a_1}+{\bf a_2})      \label{WHUU}\\
    \hat V({\bf b_1})\hat V({\bf b_2})&=&
        \hat V({\bf b_1}+{\bf b_2})     \label{WHVV}\\
    \hat U({\bf a})\hat V({\bf b})&=&
        e^{i{\bf a}\cdot{\bf b}}
            \hat V({\bf b})\hat U({\bf a})\label{WHUV}
\end{eqnarray}
corresponding to a unitary representation of the Weyl-Heisenberg
group in three dimensions.

The commutation relations \eqs{CCRxx}{CCRxp} have the
unique\footnote{More precisely, according to the Stone-von Neumann
theorem all weakly continuous, irreducible, unitary
representations of the exponentiated algebra in \eqs{WHUU}{WHUV}
are unitarily equivalent.} representation on wave functions:
\begin{eqnarray}
  (\hat x_i\psi)({\bf x})&:=&x_i\psi({\bf x})\label{Def:xi}\\
  (\hat p_j\psi)({\bf x})&:=&-i{\partial\psi\over \partial x_j}({\bf x})
                                             \label{Def:pj}
\end{eqnarray}
or, in exponentiated form,
\begin{eqnarray}
        (\hat U({\bf a})\psi)({\bf x})&:=&
            \psi({\bf x}+{\bf a})               \label{Def:U(a)}\\
        (\hat V({\bf b})\psi)({\bf x})&:=&
            e^{i{\bf b}\cdot{\bf x}}\psi({\bf x})\label{Def:V(b)}
\end{eqnarray}

The algebra generated by \eqs{CCRxx}{CCRxp} (or, equivalently,
\eqs{WHUU}{WHUV}) constitutes  the `kinematical' aspects of
constructing the quantum theory of a particle moving in three
dimensions: every system of this type has the same algebra and
representation theory. However, to this general quantisation
structure there must be added the information that specifies any
{\em particular\/} such system, which in this case means giving
the Hamiltonian (and hence the dynamical evolution). In this
context, an important requirement is that the representation of
the quantisation algebra in \eqs{CCRxx}{CCRxp} (or
\eqs{WHUU}{WHUV}) is {\em irreducible\/}. This means that the
Hamiltonian operator will necessarily be expressible as a function
of the basic canonical variables $\hat x_i,\hat p_j$, $i,j=1,2,3$.

In the {\em history\/} theory of a particle moving in three
dimensions, the quantisation algebra is \cite{ILSS98}
\begin{eqnarray}
    [\,\hat x_{it},\hat x_{js}\,]&=&0             \label{His:xtxs}\\
    {[\,}\hat p_{it},\hat p_{js}\,]&=&0            \label{His:ptps}\\
    {[\,}\hat x_{it},\hat p_{js}]&=&
                 i\delta_{ij}\delta(t-s)\hat 1    \label{His:xtps}
\end{eqnarray}
for all values of the time parameters $t,s\in\mathR$. Note that
$\hat x_{it}$ and $\hat p_{js}$ are Schr\"odinger-picture
operators, {\em not\/} Heisenberg-picture quantities---the labels
$t,s$ refer to the time at which propositions are made about the
system; they are {\em not\/} dynamical variables (these arise in a
quite different way \cite{Sav99a}).

The representation theory of the (infinite-dimensional) algebra in
\eqs{His:xtxs}{His:xtps} is much more complicated than that of the
single-time canonical algebra in \eqs{CCRxx}{CCRxp}. Once again,
the representation is required to be irreducible, so that any
operator in the theory will be a function of the history variables
$\hat x_t,\hat p_s$, with $t,s\in\mathR$.

\subsection{The Analogue for a Manifold $Q$.}
\label{SubSec:QAManQ} Now consider the problem of quantising a
system whose configuration space (or history theory analogue) is a
general finite-dimensional manifold $Q$. If $Q$ is a homogeneous
space, $G/H$, for some finite-dimensional Lie group $G$ and
subgroup $H\subset G$, the analogue of momentum is played by the
generators of $G$ which, as is expected of momentum, `translate'
around the points of $Q$. Thus $G$ appears as a finite-dimensional
subgroup of the group $\Diff{Q}$ of all diffeomorphisms of $Q$. If
$Q$ is not a homogeneous space then the entire group $\Diff{Q}$
must be used.

Thus, when $Q\simeq G/H$, we anticipate that a quantisation of
this system will include a unitary representation $g\mapsto \hat
U(g)$ of the group $G$, so that $\hat U(g_1)\hat U(g_2)=\hat
U(g_1g_2)$ for all $g_1,g_2\in G$. In the non-homogeneous case we
must use a unitary representation of $\Diff Q$, about which much
less is known.

Let us now turn to the configuration variables. When $Q$ is a
vector space, the basic such variables are the {\em linear\/} maps
from $Q$ to $\mathR$, \ie, elements of the dual $Q^*$ of $Q$. On
giving $Q$ an inner product, these are in one-to-one
correspondence with the elements of $Q$ itself. Essentially, this
is what is meant by \eq{Def:xi}, which can be read as associating
to each $3$-vector $\bf a$, an operator $\widehat{{\bf a}\cdot{\bf
x}}$ defined by $(\widehat{{\bf a}\cdot {\bf x}}\psi)({\bf
x}):={\bf a}\cdot{\bf x}\psi({\bf x})$.

For a general manifold $Q$ there are no such linear maps. However,
when $Q\simeq G/H$ it is possible to embed $Q$ as a $G$-orbit in a
vector space $W$, and the real-valued linear maps on the
lowest-dimensional such vector space then give a prefered
collection of configuration variables \cite{Isham84}.  A detailed
study leads to the following equations (cf, \eqs{WHUU}{WHUV}):
\begin{eqnarray}
        \hat U(g_1)\hat U(g_2)&=&\hat U(g_1g_2)\label{Ug1Ug2=}\\
        \hat V(\beta_1)\hat V(\beta_2)&=&
                    \hat V(\beta_1+\beta_2)\label{Vb1Vb2=}\\
        \hat U(g)\hat V(\beta)&=&
         \hat V(\beta\circ\tau_{g^{-1}})\hat U(g)\label{UgVb=}
\end{eqnarray}
for all $g,g_1,g_2\in G$ and $\beta,\beta_1,\beta_2\in W^*$.  In
\eq{UgVb=}, $\tau_g:Q\rightarrow Q$ denotes the left action of $G$
on $G/H$. Thus, if $\beta\in W^*\subset C^\infty(Q)$ then
$\beta\circ\tau_g(q):=\beta(gq)$ for all $g\in G$ and $q\in Q$.

The equations (\ref{Ug1Ug2=}) and (\ref{UgVb=}) describe an
operator representation of the semi-direct product $G\times_\tau
W^*$, with the group law
\begin{equation}
    (g_1,\beta_1)(g_2,\beta_2)=
            (g_1g_2,\beta_1+\beta_2\circ\tau_{g_1^{-1}})
            \label{GrpLawGSemiV}
\end{equation}
for all $g_1,g_2\in G$ and $\beta_1,\beta_2\in W^*$. The various
possible quantisations of the system are  deemed to be in
one-to-one correspondence with the faithful, irreducible
representations of this group.

One obvious representation of the `quantisation group'
$G\times_\tau W^*$ is given on complex-valued functions on $Q$:
\begin{eqnarray}
    (\hat U(g)\psi)(q)&:=& \psi(g^{-1}q)    \label{Def:U(g)psi=}\\
    (\hat V(\beta)\psi)(q)&:=& e^{i\beta(q)}\psi(q)
                                            \label{Def:V(beta)psi=}
\end{eqnarray}
for all $q\in Q$. However, this not the only such, and Mackey
theory \cite{Mac78} shows that the generic irreducible
representation of $G\times_\tau W^*$ is defined on functions
$\psi:Q\simeq G/H\rightarrow V$ where the vector space $V$ carries
an irreducible representation of the subgroup $H\subset G$.
Specifically:
\begin{eqnarray}
        (\hat U(g)\psi)(q)&:=& m(g,q)\psi(g^{-1}q)
                            \label{Def:U(g)Mult}\\
        (\hat V(\beta)\psi)(q)&:=& e^{i\beta(q)}\psi(q)
                            \label{Def:V(beta)Mult}
\end{eqnarray}
where the `multipliers' $m(g,q)$, $g\in G$, $q\in Q$, are linear
maps from $V$ to itself, which---in order to satisfy
\eq{Ug1Ug2=}---must satisfy
\begin{eqnarray}
            m(g_1,q)m(g_2,g_1^{-1}q)=m(g_1g_2,q)\label{mcond}
\end{eqnarray}
for all $g_1,g_2\in G$ and $q\in Q$ \cite{Kir76}.

If the manifold $Q$ is {\em not\/} a homogeneous space, the best
that can be done is to use the quantisation group $\Diff
Q\times_\tau C^\infty(Q)$ with group law (c.f., \eq{GrpLawGSemiV})
\begin{equation}
    (\phi_1,\beta_1)(\phi_2,\beta_2)=
       (\phi_1\circ \phi_2,\beta_1+\beta_2\circ\tau_{\phi_1^{-1}})
            \label{GrpLawDSemiC}
\end{equation}
for all $\phi_1,\phi_2\in\Diff Q$, $\beta_1,\beta_2\in
C^\infty(Q)$. The representation theory of this group is
considerably more complicated than that of $G\times_\tau W^*$,
although some of the representations are direct analogues of those
above.

\subsection{Quantising Models for Space or Space-time}\label{SubSec:QuCSTops} Let us turn now to the main topic of interest: namely, to construct a quantum framework for a system whose `history configuration space' $Q$ is a collection of possible space-times, such as the causal sets in Figure
\ref{Fig:Cat4objects}; or, in the canonical case, a system whose
configuration space is a specific collection of topological
spaces.

An obvious first guess for the state vectors is complex-valued
functions on $Q$, with the configuration variables---taken to be
real-valued functions $\beta$ on $Q$---being represented by the
operators $(\hat\beta\psi)(A):= \beta(A)\psi(A)$ for all $A\in Q$.
In exponentiated form, this is (c.f., \eq{Def:V(beta)psi=})
\begin{equation}
    (\hat V(\beta)\psi)(A):=
        e^{i\beta(A)}\psi(A)
\end{equation}
for all $A\in Q$.

However,  for a manifold $Q\simeq G/H$, we know that most of the
representations involve functions whose values lie in a vector
space other than $\mathC$. In addition, a causal set (and
topological space) has internal structure, and one expects this to
be represented in the quantum theory. This also suggests that
$\mathC$-valued functions on $Q$ are not sufficient as state
vectors.

What is needed is a causal-set analogue of the quantisation group
$G\times_\tau W^*$, where the generators of the Lie group $G$
physically represent momentum. Thus we seek an analogue of
momentum for a system of the type under consideration; for
example, the model in Figure \ref{Fig:Cat4objects}.

In the manifold case, each element of $G$ acts as a diffeomorphism
of $Q$. However  causal sets (and topological spaces) differ from
each other by virtue of their internal structure, and hence,
unlike a diffeomorphism, any `momentum' transformation from one
set to another is unlikely to be invertible. This suggests the use
of a {\em semi-group\/} of transformations\footnote{A semi-group
is a set $S$ equipped with a combination law that, like a group,
is associative. Inverses of elements of $S$ may not exist
however.}.

However, because different members of $Q$ have different internal
structures, it seems unlikely that a {\em single\/} semi-group
could act on {\em all\/} of $Q$. More plausible is a `local'
action in which each causal set $C\in Q$ has its own semi-group
that reflects its internal structure. In fact, the admissible
transformations from one causal set $C$ to another $C'$ will
likely depend on $C'$ as well as $C$.

We therefore arrive at the idea that to each pair of causal sets
$C,C'\in Q$ there will be a set of transformations from $C$ to
$C'$ that reflect the internal structures of $C$ and $C'$, and
which  are a local analogue of the global group $G$ of
diffeomorphisms when $Q$ is a manifold $G/H$. Clearly, the same
argument applies to a collection of topological spaces, or indeed
to any other model which involves a collection of spaces, or
space-times, with individual internal structure.

Thus, when quantising a system whose configuration space (or
history equivalent) is a collection $Q$ of sets with structure,
the starting point will be an association to each pair $A,A'\in Q$
of a set $T(A,A')$ of transformations from $A$ to $A'$ that, in
some suitable sense, `respect', or `reflect', the internal
structures of $A$ and $A'$. From a physical perspective, these
transformations will be related in some way to an analogue of
momentum.

In the case of causal sets, the natural choice for the
transformations from $A$ to $A'$ is {\em
order-preserving\/}\footnote{A map $\alpha:A\rightarrow A'$ is
{\em order-preserving} if whenever $a,b\in A$ are such that $a\leq
b$, then $\alpha(a)\leq \alpha(b)$.} maps from $A$ to $A'$; for
topological spaces, the natural choice would be {\em continuous\/}
maps from $A$ to $A'$.

Note that transformation sets of the type postulated above arise
also in the case of a group $G$ acting on a manifold $Q$, albeit
in a somewhat different way since, in this case, there is no
internal structure to respect. Specifically, we define an
`admissible transformation' from $q\in Q$ to $q'\in Q$ to be any
group element $g\in G$ such that $q'=gq$; thus
\begin{equation}
            T(q,q')=\{g\in G\mid q'=gq\}.    \label{Def:Tqq'}
\end{equation}

The following question arises in general. Namely, if $A,A',A''$
are three elements in $Q$, are there any relations between the
sets $T(A,A')$, $T(A',A'')$ and $T(A,A'')$? The obvious
requirement is that if $\alpha \in T(A,A')$, and $\beta\in
T(A',A'')$ then the composition map $\beta\circ \alpha$ (defined
by $\beta\circ\alpha(a):=\beta(\alpha(a))$ for all $a\in A$)
should belong to $T(A,A'')$, \ie, it should respect the internal
structure. This works in the example of causal sets, since the
combination of two order-preserving maps is itself order
preserving. Similarly, in the case of topological spaces, the
combination of two continuous maps is continuous.

Note that a natural `composition' also exists in the case of a
group $G$ acting on a manifold $Q$, with the `transformations'
from $q$ to $q'$ being defined as in \eq{Def:Tqq'}. For if
$q'=g_1q$ and $q''=g_2q'$ then $q''=(g_2g_1)q$, and hence this
case fits into the general scheme if the composition $g_2\circ
g_1$ is defined to be the group product $g_2g_1$.

This example suggests another property for the sets of
transformations $T(A,A')$, $A,A'\in Q$. Namely: a group product
law is {\em associative\/}, so that $g_1(g_2g_3)=(g_1g_2)g_3$ for
all $g_1,g_2,g_3\in G$; and this suggests that we should require
the same for the general sets $T(A,A')$. Thus if $\alpha\in
T(A,A')$, $\beta\in T(A',A'')$ and $\gamma\in T(A'',A''')$ then $
\gamma\circ (\beta\circ\alpha)= (\gamma\circ\beta)\circ\alpha$. In
fact, this is automatically true for the composition of functions,
and hence for the examples of causal sets and topological spaces.

Similarly, it is natural to require the existence of a unit
element $\id_A$ in each $T(A,A)$ so that if $\alpha\in T(A',A)$
then $\id_{A'}\circ\alpha=\alpha=\alpha\circ\id_A$.  In examples
like causal sets, topological spaces, etc, these conditions are
satisfied if we choose $\id_A$ to be the identity map from the set
$A$ to itself. Note that the existence of these unit elements
means that each set $T(A,A)$, $A\in Q$, is a {\em monoid\/}, \ie,
a semi-group with a unit element.

In summary, for any collection $Q$ of entities with internal
structure, the quantisation framework will involve an association
to each pair $A,A'$ in $Q$ of a set of transformations $T(A,A')$
such that:
\begin{enumerate}
\item[i)] If $\alpha\in T(A,A')$ and $\beta\in T(A',A'')$ then a
combination $\beta\circ\alpha$ can be defined which belongs to
$T(A,A'')$.

\item[ii)] To each $A\in Q$ there is a unit element $\id_A\in
T(A,A)$.

{\item[iii)] If $\alpha\in T(A,A')$, $\beta\in T(A',A'')$ and
$\gamma\in T(A'',A''')$ then
\begin{equation}\gamma\circ (\beta\circ\alpha)=
        (\gamma\circ\beta)\circ\alpha.\label{Assoc}
\end{equation}}
\end{enumerate}

However, these are nothing but the axioms for a {\em category\/}
provided that the `transformations' in $T(A,A^\prime)$ are
interpreted as arrows! In fact, the motivating examples of causal
sets and topological spaces are special types of category in which
the objects are {\em sets\/} with structure, and the arrows are
structure-preserving maps between these sets. For example, Figure
\ref{Fig:Cat4objects} represents a category with four objects, in
which the arrows/morphisms between a pair of causal sets $C_i,C_j$
are defined to be the order-preserving maps from $C_i$ to $C_j$.
Similarly, a collection of topological spaces is a category in
which the arrows between a pair of topological spaces are defined
to be the continuous maps between them.

However, not all categories are collections of structured sets,
and we may wish to include this type of category too. Indeed, as
shown above, the familiar example of a configuration space
manifold $Q$ on which a group $G$ acts, can  be realised in this
category sense. Namely, the objects in the category are the points
of $Q$, and the arrows from $q$ to $q'$ are defined to be the
group elements $g$ such that $q'=gq$. Note that the set
$T(q,q)=\{g\in G\mid gq=q\}$ is the `little group', or isotropy
group at the point $q\in Q$. Thus, in this case, the monoid
$T(q,q)$, $q\in Q$, is actually a group.

Thus we have naturally arrived at the problem of finding the
quantum analogue of a system for which (i) the configuration space
(or history-theory analogue) is the set, $\Ob\Q$, of all {\em
objects\/} in some category\footnote{Strictly speaking, the
category has to be `small', by which is meant that the collection
of objects and arrows are genuine {\em sets\/}, not higher-order
classes. For example, the category of all sets is not small, since
the collection of all sets is not itself a set.} $\Q$; and (ii)
the sets $T(A,A')$ are identified with the sets, $\Hom{A,A'}$, of
all {\em arrows\/} (or {\em morphisms}) in the category with
domain $A$ and range $A'$. We expect that quantising this system
will involve representing the elements of $\Hom{A,A'}$ with
operators on a Hilbert space in some way.

This use of category theory may sound rather abstract but, as I
hope is clear from the examples above, it actually arises quite
naturally. In particular, if we can find a way of associating a
quantum structure with {\em any\/} (small) category $\Q$, then
this will include the known quantisation scheme for homogeneous
manifolds $Q\simeq G/H$, but will extend it to a much wider class
of physical system.

Note that it is possible to have two categories with the same set
of objects but different arrows. This has immediate physical
implications since, if we follow the analogy of $Q\simeq G/H$, we
will be seeking an operator representation of configuration
functions (\ie, functions on objects) and arrows that is
irreducible, and hence such that any other operator in the
theory---for example, the Hamiltonian, or (in a history theory)
the decoherence operator---can be written as functions of these
basic operators. For example, in Figure \ref{Fig:Cat4objects}the
arrows from one causal set $C_i$ to another $C_j$ could be chosen
to be any order-preserving function from $C_i$ to $C_j$; but we
could also restrict ourselves to order-preserving functions that
are one-to-one, \ie, which {\em embed\/} $C_i$ in $C_j$. In
practice, a lot of physical input will be required to choose the
objects and arrows in any particular category.

\section{The Theory of Arrow Fields}
\label{Sec:ArrowFields} As emphasised in Section
\ref{SubSec:QuCSTops}, when quantising on a category $\Q$ we
expect to assign operators to arrows in such a way as to represent
the arrow combination law.

However, this runs into an immediate problem. For suppose $\hat
d(f)$, $f\in\Hom\Q$, are such operators. Then, representing the
arrow combination law presumably means
\begin{equation}
        \hat d(g)\hat d(f)=\hat d(g\circ f) \label{dgdf=dgf}
\end{equation}
whenever the arrows $f,g$ are such that $\Ran f=\Dom g$, \ie, such
that the combination $g\circ f$ is well defined. Then
\eq{dgdf=dgf} is to be viewed as the category analogue of the
representation $\hat U(g_1)\hat U(g_2)=\hat U(g_1g_2)$ of the Lie
group $G$ in the quantisation algebra \eqs{WHUU}{WHUV} for a
system whose configuration space is a manifold $Q\simeq G/H$. But
the problem with \eq{dgdf=dgf} is that the right hand side is only
defined when the combination $g\circ f$  exists, whereas the left
hand side is defined for {\em all\/} arrows
$f,g$.\footnote{Equivalently, the collection of arrows in a
category is a {\em partial\/} semigroup under the law of arrow
combination, whereas the (bounded) operators on a Hilbert space
are a (full) semigroup.}

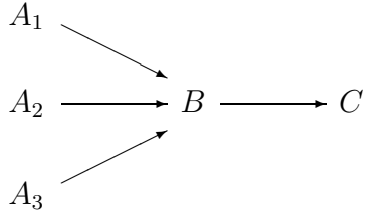
\begin{figure}[h]
\begin{picture}(100,100)(-40,-20)
        \put(80,27){$A_2$}
        \put(100,30){\vector(1,0){40}}
        \put(145,27){$B$}
        \put(160,30){\vector(1,0){40}}
        \put(205,27){$C$}
        \put(100,60){\vector(2,-1){40}}
        \put(100,0){\vector(2,1){40}}
        \put(80,60){$A_1$}
        \put(80,-8){$A_3$}
\end{picture}
\caption[]{An arrow field in a category with 5
objects}\label{Fig:AF5Ob}
\end{figure}

One possibility would be to define $\hat d(g\circ f):=0$ if $\Ran
f\neq\Dom g$. However, we shall proceed in a somewhat different
way that involves introducing the idea of an `arrow field'. An
{\em arrow field\/}\footnote{The concept of an arrow field has
been introduced independently by Gilbert \cite{Gil03} in a
different context, where he calls it a `flow' on the category. He
refers back in turn to an unpublished preprint of Chase
\cite{Cha77}. I am very grateful to Nick Gilbert for drawing my
attention to this work.} $X$ is defined to be an assignment to
each object $A$ of an arrow $X(A)$ whose domain is $A$. For
example, consider a category with five objects $A_1,A_2,A_3,B,C$,
and with the arrow field shown in Figure \ref{Fig:AF5Ob}.
Thus\footnote{The notation $f:A\rightarrow B$ means that $f$ is an
arrow with domain $A$ and range $B$. In categories where the
objects are structured sets, such an arrow will be a
(structure-preserving) map from the set $A$ to the set $B$.}
$X(A_1):A_1\rightarrow B$, $X(A_2):A_2\rightarrow B$,
$X(A_3):A_3\rightarrow B$, $X(B):B\rightarrow C$, and $X(C)$ (not
shown) is the identity arrow $\id_C:C\rightarrow C$.

The reason for introducing arrow fields is that, unlike arrows,
they {\em can\/} always be combined. More precisely, if $X_1$ and
$X_2$ are arrow fields, we define a new arrow field,
denoted\footnote{Note that the analogous quantity in
\cite{IshQCT1_03}, \cite{IshQCT2_03}, \cite{IshQCT3_03} was
denoted $X_2\& X_1$, rather than $X_1\&X_2$. Of course, this is
entirely a matter of convention, but the choice made in the
present paper is more convenient for certain purposes.}
$X_1\&X_2$, by
\begin{equation}
        (X_1\&X_2)(A):=X_2(\Ran{X_1(A)})\circ X_1(A)
        \label{Def:X2X1=}
\end{equation}
for all objects $A$. More pictorially, if $A\smapright{X_1(A)}
B\smapright{X_2(B)}C$ then $(X_1\&X_2)(A)$ is the arrow from $A$
to $C$ defined by $(X_1\&X_2)(A):=X_2(B)\circ X_1(A)$.

An important feature of this combination law is that it is {\em
associative}. Thus, for any three arrow fields $X_1,X_2,X_3$ we
have
\begin{equation}
         X_1\&(X_2\& X_3)=(X_1\&X_2)\&X_3\label{AssocAF}.
\end{equation}
Furthermore, if an arrow-field $\iota$ is defined by
\begin{equation}
        \iota(A):=\id_A                 \label{Def:iota}
\end{equation}
for all objects $A$, then $\iota$ is an identity element for the
`$\&$' composition law, \ie, $\iota\& X=X\&\iota =X$ for all arrow
fields $X$.

It follows that the set of all arrow fields on the category $\Q$
is a {\em monoid\/} when equipped with the combination law in
\eq{Def:X2X1=} and the identity in \eq{Def:iota}. We will denote
this monoid of arrow fields by $\AF\Q$. A key idea in constructing
a quantum theory on $\Q$ is that the monoid $\AF\Q$ can play a
role analogous to that of $\Diff{Q}$ in the case of a manifold
$Q$.

For a manifold $Q\simeq G/H$, the group $G$ acts on $Q$ as a group
of transformations, and we need the analogue of that here.
Specifically, we define an action of an arrow field $X$ on the set
$\Ob\Q$ by
\begin{equation}
        \rho_X(A):=\Ran{X(A)}   \label{Def:rhoXA}
\end{equation}
for all objects $A$; note that this is  a {\em right\/} action
since
\begin{equation}
    \rho_{X_2}\circ\rho_{X_1}=\rho_{X_1\&X_2}
\end{equation}
for all arrow fields $X_1,X_2$. An example, of this action is
provided by the arrow field shown in Figure \ref{Fig:AF5Ob}: the
objects $A_1,A_2,A_3$ are mapped to $B$, the object $B$ is mapped
to $C$, and $C$ is mapped to itself.

Individual arrows are associated with special arrow fields. Thus,
if $f:A\rightarrow B$ we define an arrow field $X_f$ by
\begin{equation}
    X_f(C):=\left\{ \begin{array}{ll}
       \mbox{$f$ \ \ if $C=A$;} \label{Def:Xf}\\
           \mbox{$\id_C$ otherwise.}
                 \end{array}
        \right.
\end{equation}
This arrow field acts on $\Ob\Q$ by $\rho_{X_f}(A)=B$, and
$\rho_{X_f}(C)=C$ for all objects $C\neq A$.

Note that if $f$ and $g$ are arrows in the monoid $\Hom{A,A}$,
then
\begin{equation}
    X_f\&X_g=X_{g\circ f}       \label{XfXg=}
\end{equation}
and hence the set of all arrow fields of the type $X_f$,
$f\in\Hom{A,A}$, gives a (anti)-representation of the monoid
$\Hom{A,A}$ in the monoid $\AF\Q$.

\section{Arrow-Field Quantum Theory}
\label{Sec:ArrowFieldQT}
\subsection{The Basic Quantum Algebra}
The next step is to find an analogue for a category $\Q$ of the
quantisation group \eqs{Ug1Ug2=}{UgVb=} for a configuration
manifold $Q$. The construction of this group involves the
cotangent bundle $T^*Q$, which is the classical state space of the
system. However, there is no immediate analogue of this for a
general category, and therefore we shall proceed in a more
heuristic way.

For a general manifold $Q$, physical momenta are associated with
the diffeomorphism group $\Diff Q$, and---as argued already---the
analogue of this for a category $\Q$ is the monoid $\AF\Q$ which
acts on the `configuration space' $\Ob\Q$ according to
\eq{Def:rhoXA}. The configuration variables on a manifold are
smooth, real-valued functions on $Q$, and the analogue of this for
a category would be some `appropriate'  subspace of the vector
space $F(\Ob\Q,\mathR)$ of all real-valued functions on $\Ob\Q$.
However, for a general category there is no obvious preferred such
subspace, which obliges us to work with the entire space
$F(\Ob\Q,\mathR)$, albeit with the understanding that for specific
categories $\Q$ there may be natural analogues of the subspace
$C^\infty(Q)\subset F(Q,\mathR)$ (or even, perhaps, of the
subspace $W\subset C^\infty(Q)$ for the case $Q\simeq G/H$).

The quantisation group for a general manifold $Q$ is the
semi-direct product $\Diff{Q}\times_\tau C^\infty(Q)$ with the
group law in \eq{GrpLawDSemiC}. This suggests strongly that the
analogue for a category $\Q$ should be the semi-direct product,
$\AF\Q\times_\rho F(\Ob\Q,\mathR)$, of the monoid $\AF\Q$ and the
vector space $F(\Ob\Q,\mathR)$. The combination law is
\begin{equation}
    (X_1,\beta_1)(X_2,\beta_2)=
            (X_1\&X_2,\beta_1+\beta_2\circ\rho_{X_1})
            \label{GrpLawCQM}
\end{equation}
for all $X_1,X_2\in\AF\Q$ and $\beta_1,\beta_2\in
F(\Ob\Q,\mathR)$. We shall refer to $\AF\Q\times_\rho
F(\Ob\Q,\mathR)$ as the {\em category quantisation monoid\/} (or
just CQM) for the category $\Q$.

\subsection{Representations on `wave-functions'}
Although plausible, the derivation above of the CQM may seem a
little ad hoc, and it is useful therefore to check the overall
consistency of these ideas by looking for the simplest type of
representation, which involves taking state vectors to be
functions $\psi:\Ob\Q\rightarrow\mathC$.

Motivated by \eq{Def:U(g)psi=}, we define, for each arrow field
$X$, the operator
\begin{equation}
    (\hat a(X)\psi)(A):=\psi(\rho_XA)=\psi(\Ran{X(A)}).
    \label{Def:a(X)psiWFn}
\end{equation}
It can readily be checked that these operators satisfy
\begin{equation}
        \hat a(X_1)\hat a(X_2)=\hat a(X_1\& X_2)
                        \label{aX1aX2=}
\end{equation}
for all arrow fields $X_1,X_2$. Thus we have a representation of
the monoid of arrow fields. Equation (\ref{aX1aX2=}) is a precise
analogue of the relation \eq{Ug1Ug2=} for a manifold.

Turning now to the configuration variables, and guided by
\eq{Def:V(beta)psi=}, we define, for each $\beta\in
F(\Ob\Q,\mathR)$,  the operator $\hat\beta$:
\begin{equation}
    (\hat\beta\psi)(A):=\beta(A)\psi(A)
\end{equation}
and its exponentiated form:
\begin{equation}
        (\hat V(\beta)\psi)(A):=e^{i\beta(A)}\psi(A)
                                    \label{Def:V(beta)psiWFn}
\end{equation}
for all objects $A$ in the category.

These operators satisfy the relations (c.f., \eqs{Ug1Ug2=}{UgVb=})
\begin{eqnarray}
    \hat a(X_1)\hat a(X_2)&=&\hat a(X_1\& X_2)  \label{CQMaa}\\
    \hat V(\beta_1)\hat V(\beta_2)&=&\hat V(\beta_1+\beta_2)
                                                \label{CQMVV}\\
    \hat a(X)\hat V(\beta)&=&\hat V(\beta\circ\rho_X)\hat a(X)
                                                \label{CQMaV}
\end{eqnarray}
for all arrow fields $X,X_1,X_2$ and real-valued functions
$\beta,\beta_1,\beta_2$. The relations in \eqs{CQMaa}{CQMaV}
constitute a representation of the CQM, $\AF\Q\times_\rho
F(\Ob\Q,\mathR)$, defined in \eq{GrpLawCQM}. This helps to justify
the claim that the CQM is the category analogue of the group
$\Diff{Q}\times_\tau C^\infty(Q)$ used when the configuration
space is a manifold $Q$.

Thus I shall define the {\em quantisations on the category $\Q$\/}
to be in one-to-one correspondence with faithful, irreducible
representations of the category quantisation monoid
$\AF\Q\times_\rho F(\Ob\Q,\mathR)$. However, as we shall see, the
representation on wave-functions given by \eq{Def:a(X)psiWFn} and
\eq{Def:V(beta)psiWFn} is not faithful, and therefore we need some
more complicated representations of the CQM.

In the case of a configuration manifold $Q$, the operators $\hat
U(g)$, $g\in G$, and $\hat V(\beta)$, $\beta\in C^\infty(Q)$, are
required to be {\em unitary\/}, and we may wonder what the
analogue of that would/should be for the CQM. This requires
placing an inner product on the quantum states and, in the simple
example of the `wave-function' representation given by
\eq{Def:a(X)psiWFn} and \eq{Def:V(beta)psiWFn}, this would be of
the form
\begin{equation}
        \braket\psi\phi:=\int_{\Ob\Q}d\mu(A)\,\psi(A)^*\phi(A)
                    \label{Def:braket_psi_phiWFns}
\end{equation}
where $\mu$ is some suitable measure on the space $\Ob\Q$. Of
course, at this stage it is not clear what `suitable' means since
representations of a monoid (as opposed to a group) are not
unitary (because of the absence of inverse elements) whereas, in
the group case, it is the desired unitarity of the representations
that essentially determines the measure.

However, if the category has only a finite number of objects (such
as the toy model in Figure \ref{Fig:Cat4objects}) it is natural to
use the simple `counting' measure
\begin{equation}
   \braket\psi\phi:=\sum_{A\in\Ob\Q}\psi(A)^*\phi(A).
                \label{Def:braketpsiphi}
\end{equation}
With due care, this can also be used if the category has a
countably infinite number of objects.

\subsection{Introducing a Multiplier}
An important requirement for a representation of the CQM is that
it {\em separates\/} arrows. Namely: if $f,g$ are different arrows
with the same domain and the same range, then $\hat a(X_f)\neq
\hat a(X_g)$. In particular, this applies if the domain and range
of the arrows $f,g$ are a single object $A$, in which case the
requirement is that the (anti-) representation of each monoid
$\Hom{A,A}$, $A\in\Ob\Q$ given by the operators $\hat a(X_f)$,
$f\in\Hom{A,A}$, (using \eq{XfXg=}) is faithful. Indeed, the
monoid $\Hom{A,A}$ is one measure of the internal structure of the
object $A$, and we want the quantum  theory to reflect this
structure in a faithful way.

However,  the representation given by the operators $\hat a(X)$
defined in \eq{Def:a(X)psiWFn} is clearly far from faithful.
Indeed, since the action of $\hat a(X)$ on a state function at an
object $A$ depends only on the object $\Ran{X(A)}$, arrows with
the same domain and range are {\em never\/} separated. In
particular, the monoid $\Hom{A,A}$ is represented trivially for
all objects $A$.

Turning for guidance to the case of a configuration manifold
$Q\simeq G/H$, the theory of induced representations shows that
the most general irreducible representation of the quantisation
group $G\times_\tau W^*$ is specified by (i) an orbit of $G$ on
$W$; and (ii) an irreducible representation of $H$ on a Hilbert
space $V$. This representation of $G\times_\tau W^*$ is defined on
cross-sections of a vector bundle over the orbit with fibre $V$.
The generic orbits on $W$ are diffeomorphic to the configuration
manifold $Q$.

This suggests that in the category case, we should introduce some
bundle $\K$ of Hilbert spaces over the set of objects $\Ob\Q$,
with an associated inner product on the sections of this bundle
given by
\begin{equation}
\braket\psi\phi:=
    \int_{\Ob\Q}d\mu(A)\,\langle\psi(A),\phi(A)\rangle_{\K(A)}
                    \label{Def:braket_psi_phiVFns}
\end{equation}
where $\langle\psi(A),\phi(A)\rangle_{\K(A)}$ denotes the inner
product in the Hilbert space $\K(A)$ of the vectors
$\psi(A),\phi(A)\in\K(A)$, where $\K(A)$ is the fibre over the
object $A$.

A necessary requirement for this Hilbert bundle to give an
arrow-separating representation is that, for each object $A$, the
vector space $\K(A)$ should carry a faithful, irreducible,
representation of the monoid $\Hom{A,A}$ of arrows from $A$ to
itself. However, the internal structure of an object $A$, and
hence of $\Hom{A,A}$, generally depends on the object $A$ (for
example, consider the monoid of order-preserving maps from each
object in Figure \ref{Fig:Cat4objects} to itself), and this means
that the Hilbert space $\K(A)$ should depend on $A$. But in a
normal vector bundle the fibres are all isomorphic to each other,
and hence we are looking for something more general: as we shall
see shortly, this is a {\em presheaf\/} of Hilbert spaces.

With this caveat, we now consider the category analogue of the
multiplier group representations defined in
\eqs{Def:U(g)Mult}{Def:V(beta)Mult}. Thus, to each arrow field $X$
and object $A$, we require a linear map (a `multiplier')
$m(X,A):\K(\rho_XA)\rightarrow \K(A)$ from the Hilbert space
$\K(\rho_XA)$ to the Hilbert space $\K(A)$. Then, in direct
analogy with \eqs{Def:U(g)Mult}{Def:V(beta)Mult}, we define the
operators:
\begin{eqnarray}
    (\hat a(X)\psi)(A)&:=& m(X,A)\psi(\rho_XA)
                                    \label{Def:a(X)PS}\\
    (\hat V(\beta)\psi)(A)&:=&e^{i\beta(A)}\psi(A).
                                    \label{Def:V(beta)PS}
\end{eqnarray}

It is easy to check that \eq{Def:a(X)PS} gives a representation of
the monoid $\AF\Q$ provided the multipliers satisfy the conditions
\begin{equation}
        m(X_1,A)m(X_2,\rho_{X_1}A)=m(X_1\& X_2,A)   \label{mcondPS}
\end{equation}
for all arrow fields $X_1,X_2$ and objects $A$. This is the
precise analogue for $\AF\Q$ of the well-known condition
\eq{mcond} on the multipliers in a representation of a group $G$
that acts on a configuration space $Q$.

\subsection{The Presheaf Perspective}
As things stand, the multiplier $m(X,A):\K(\rho_XA)\rightarrow
\K(A)$ could depend on the value of the arrow field $X$ on objects
other than $A$. However, it seems natural to try imposing the
condition that $m(X,A)$ depends only on the arrow $X(A)$. Thus we
suppose that, for each arrow $f:A\rightarrow B$, there is a linear
map $\kappa(f):\K(B)\rightarrow \K(A)$ such that
$m(X,A)=\kappa(X(A))$.

The requirement on the multipliers in \eq{mcondPS} then becomes
the condition
\begin{equation}
        \kappa(f)\kappa(g)=\kappa(g\circ f) \label{kcondPS}
\end{equation}
for all arrows $f,g$ such that $\Ran f=\Dom g$. This is
illustrated in the diagram
\begin{equation}
    \bundle{\K(A)}{}{A}\bundlemapLR{\kappa(f)}{f}
    \bundle{\K(B)}{}{B}\bundlemapLR{\kappa(g)}{g}
    \bundle{\K(C)}{}{C}
\end{equation}

However, these are precisely the conditions for a {\em presheaf\/}
of Hilbert spaces over the category $\Q$. Thus a representation of
the CQM can be obtained from each such presheaf of Hilbert spaces
over $\Q$.\footnote{Something rather similar happens
mathematically in the case of topological quantum field theory,
and the category quantisation methods may have important
applications to that area. This is currently being investigated.}

\subsection{The adjoints of the operators $\hat a(X)$}
As mentioned earlier, because of the existence of non-invertible
elements, a monoid representation will not be unitary. This lends
interest to the question of what the adjoints of the operators
$\hat a(X)$ look like.

The main features can be illustrated with the simple example of a
category whose number of objects is finite, and with state
functions $\psi:\Ob\Q\rightarrow\mathC$ (\ie, no multipliers), as
in \eq{Def:a(X)psiWFn} and \eq{Def:V(beta)psiWFn}. The simplest
inner product is \eq{Def:braketpsiphi}, and using standard Dirac
notation, with $\psi(A)$denoted $\braket{A}\psi$, we have
\begin{eqnarray}
   \hat a(X)^\dagger\ket{B}&=&\ket{\rho_X B}=\ket{\Ran{X(B)}}
                        \label{a(X)daggerKetB}\\[2pt]
        \hat a(X)\ket{B}&=&\sum_{A\in\rho^{-1}_X\{B\}}\ket{A}
                    \label{a(X)KetB}
\end{eqnarray}
where $\rho_X^{-1}\{B\}$ denotes the set of all objects $A$ such
that $\rho_X(A)=B$, \ie, such that the range of the arrow $X(A)$
is $B$.

For example, for the arrow field illustrated in Figure
\ref{Fig:AF5Ob} in a simple $5$-object category, we have
(remembering that $X(C)=\id_C$)
\begin{eqnarray}
    \hat a(X)^\dagger\ket{A_1}&=&\hat a(X)^\dagger\ket{A_2}=
                \hat a(X)^\dagger\ket{A_3}=\ket{B}
                                        \label{a(X)daggerketA5cat}\\
    \hat a(X)^\dagger\ket{B}&=&\ket{C}  \label{a(X)daggerketB5cat}\\
    \hat a(X)^\dagger\ket{C}&=&\ket{C}  \label{a(X)daggerketC5cat}
\end{eqnarray}
and
\begin{eqnarray}
    \hat a(X)\ket{A_1}&=&\hat a(X)\ket{A_2}=\hat a(X)\ket{A_3}=0
                            \label{a(X)ketA5cat}\\
    \hat a(X)\ket{B}&=&\ket{A_1}+\ket{A_2}+\ket{A_3}
                            \label{a(X)ketB5cat}\\
    \hat a(X)\ket{C}&=&\ket{B}+\ket{C}
                            \label{a(X)ketC5cat}
\end{eqnarray}

Note that the right hand side of \eq{a(X)ketA5cat} is $0$ since no
objects in the category are mapped to $A_1$, $A_2$, or $A_3$ by
the arrow field $X$. In that sense, the operator $\hat a(X)$
resembles an {\em annihilation\/} operator.

On the other hand, we see from \eq{a(X)daggerKetB} that there are
no states $\ket{B}$ such that $\hat a(X)^\dagger\ket{B}=0$ (since
every arrow has a range), and in that limited sense $\hat
a(X)^\dagger$ resembles a {\em creation\/} operator. Of course,
\eq{a(X)daggerKetB} does not mean that there are {\em no\/} states
$\ket\psi$ such that $\hat a(X)^\dagger\ket\psi=0$. For example,
for the arrow field represented by Figure \ref{Fig:AF5Ob} we have
$\hat a(X)^\dagger\ket{A_1}=\ket{B}$ and $\hat
a(X)^\dagger\ket{A_2}=\ket{B}$, and hence $\hat
a(X)^\dagger(\ket{A_1}-\ket{A_2})=0$.

In a similar way, we find
\begin{eqnarray}
    \hat a(X)\hat a(X)^\dagger\ket{A}&=&
        \sum_{C\in\rho_X^{-1}\{\rho_XA\}}\ket{C}
                        \label{a(X)a(x)daggerKetA=}\\
    \hat a(X)^\dagger\hat a(X)\ket{A}&=&
            |\rho_X^{-1}\{A\}|\ket{A}\label{a(X)daggera(x)KetA=}
\end{eqnarray}
where $|\rho_X^{-1}\{A\}|$ denotes the number of elements in the
set $\rho_X^{-1}\{A\}$. For example, for the arrow field in Figure
\ref{Fig:AF5Ob} we have
\begin{equation}
        \hat a(X)^\dagger \hat a(X)\ket{B}=3\ket{B}.
\end{equation}

The results for multiplier representations are similar to the
above; for details see \cite{IshQCT1_03}.

\subsection{The analogue of momentum}
If $Q$ is a manifold $G/H$, the analogue of momentum is played by
the elements of the Lie algebra, $L(G)$, of the Lie group $G$.
Thus, to each $k\in L(G)$ there is associated the one-parameter
subgroup $t\mapsto \exp{itk}$ of $G$, and this is represented on
the quantum Hilbert space by a one-parameter group of unitary
operators. Then, by Stone's theorem, there exists a unique
self-adjoint operator $\hat k$ such that $\hat
U(\exp{itk})=e^{it\hat k}$.

The operators $\hat k$, $k\in L(G)$, give a Hilbert space
representation of the Lie algebra $L(G)$. From a physical
perspective, these operators are the analogue for this system of
momentum variables.

The situation for a general category $\Q$ is different. The
analogue of the operators $\hat U(g)$, $g\in G$, is the operators
$\hat a(X)$, $X\in\AF\Q$. But there are some important
differences: (i) there is no analogue of the Lie algebra of $G$;
and (ii) the operators $\hat a(X)$ are not unitary. However,
quantum physical variables are associated with self-adjoint
operators, and we have to construct these in some way from what we
have: namely the operators $\hat a(X)$, $X\in\AF\Q$. To this end
we define
\begin{eqnarray}
    \hat \alpha(X)&:=&
        {1\over 2}\Big(\hat a(X)+\hat a(X)^\dagger\Big)
                            \label{Def:hatalpha}\\
    \hat\beta(X) &:=&
        {1\over 2i}\Big(\hat a(X)-\hat a(X)^\dagger\Big)
                            \label{Def:hatbeta}
\end{eqnarray}
so that $\hat a(X)=\hat\alpha(X)+i\hat\beta(X)$, where
$\hat\alpha(X)$ and $\hat\beta(X)$ are self-adjoint.

For a manifold $Q\simeq G/H$, we can write $\exp {it\hat
k}=\cos{t\hat k}+i\sin{t\hat k}$, and hence $\hat\alpha(X)$ and
$\hat\beta(X)$ are analogous to $\cos{t\hat k}$ and $\sin{t\hat
k}$ respectively. The expansion $\sin r\simeq r+O(r^3)$ then
suggests that $\hat\beta(X)$ is an analogue of a momentum
operator.

To clarify this, consider a model quantum theory with basic states
$\ket{n}$ where $n=0,1,2,\ldots$, and with the translation
operator $\hat T^\dagger$ defined by\footnote{This model can be
viewed as a special case of the category scheme in which the
objects are the integers $0,1,2\ldots$, and there is a single
arrow from $n$ to $m$ if $m\geq n$. The operator $\hat T$ then
corresponds to the special arrow field that associates to each $n$
the arrow $n\rightarrow n+1$.}
\begin{equation}
    \hat T^\dagger\ket{n}:=\ket{n+1}\label{Def:TDaggerKetn}
\end{equation}
for all $n$. This is the analogue of \eq{a(X)daggerKetB}, and the
analogue of \eq{a(X)KetB} is
\begin{equation}
\hat T\ket{n}:=\left\{ \begin{array}{ll}
            \ket{n-1}&\mbox{\ if $n>0$;}
                                    \\
                 \ 0 & \mbox{\ if $n=0$.}
                 \end{array}\label{Def:TKetn}
        \right.
\end{equation}
Then, the operator $\hat\beta:=(\hat T-\hat T^\dagger)/2i$ acts on
the states $\ket{n}$ as the difference operator
\begin{equation}
    \hat\beta\ket{n}={1\over 2i}\Big(\ket{n+1}-\ket{n-1}\Big)
\end{equation}
for all $n>0$ (and with $\hat\beta\ket{0}={i\over 2}\ket{1}$).
This can be viewed as a discrete analogue of the familiar momentum
differential operator.

In the category case, \eqs{a(X)daggerKetB}{a(X)KetB} give
\begin{equation}
    \hat\beta(X)\ket{B}={i\over 2}\Big(\ket{\Ran{X(B)}} -
          \sum_{A\in\rho^{-1}_X\{B\}}\ket{A}\Big)
\end{equation}
which is also a type of difference operator. So, in that sense,
$\hat\beta(X)$ is the category equivalent of momentum.

Similarly, if we define $\hat\alpha:=(\hat T+\hat T^\dagger)/2$,
then (if $n>0$)
\begin{equation}
    \hat\alpha\ket{n}={1\over 2}\Big(\ket{n-1}+\ket{n+1}\Big)
\end{equation}
which is a type of `average position' operator. In the category
case we have
\begin{equation}
    \hat\alpha(X)\ket{B}={1\over 2}\Big(
          \sum_{A\in\rho^{-1}_X\{B\}}\ket{A}+
                    \ket{\Ran{X(B)}}\Big)
\end{equation}
which is therefore a type of `average configuration point'
operator.

We note that the special arrow fields of the type $X_f$ defined in
\eq{Def:Xf} satisfy $X_f\&X_f=X_f$. As a result, if we define
$\hat a(f):=\hat a(X_f)$ we find that $\hat a(f)\hat a(f)=\hat
a(f)$, so that $\hat a(f)$ is a non-hermitian projection operator.
It is easy to check that, in this case, we have, for all arrows
$f$,
\begin{equation}
    \hat\alpha(f)^2-\hat\alpha(f)=\hat\beta(f)
\end{equation}
and hence $\hat\beta(f)$ is not an independent variable from
$\hat\alpha(f)$.

\subsection{Irreducibility of the representations}
Let us now say something about the irreducibility, or otherwise,
of the multiplier representations of the category quantisation
monoid.

It seems unlikely that one could develop a complete representation
theory of the CQM for an arbitrary small category $\Q$. However,
in the manifold analogy with  $Q\simeq G/H$, it is important that
$G$ acts {\em transitively\/} on $Q$. If this is not the case---so
that $Q$ can be decomposed into more than one $G$-orbit---then
there is a corresponding decomposition of the group representation
of $G\times_\tau W$ into a direct sum or direct integral. It seems
plausible that this has a category analogue, and so a natural
question is whether $\Ob\Q$ is a single orbit under the action of
$\AF\Q$.

Because of the absence of inverse elements, the concept of an
`orbit' is more subtle for an action of a monoid on a set than it
is for a group. However, on looking at the operators $\hat a(X)$
and $\hat a(X)^\dagger$---as, for example, in the simple
expressions in \eq{a(X)KetB} and \eq{a(X)daggerKetB}---it seems
natural to define a subset $O$ of $\Ob\Q$ to be `connected' if for
any pair of objects $A,B\in O$ there exists a finite collection of
objects $\{A_1,A_2,\ldots,A_N\}\subset O$, with $A_1=A$, $A_N=B$
and such that, for all $i=1,2,\ldots N-1$, there exists an arrow
with domain $A_i$ and range $A_{i+1}$, or an arrow with range
$A_i$ and domain $A_{i+1}$.

Clearly, if $\Ob\Q$ decomposes into a disjoint union of connected
subsets, then the representation of the CQM will decompose in a
corresponding way. Thus a necessary condition for the
representation to be irreducible is that $\Ob\Q$ is connected.
However, connectedness alone is not sufficient to guarantee
irreducibility. Another necessary requirement is that, for all
objects $A$, the representations of $\Hom{A,A}$ on the Hilbert
spaces $\K(A)$ should be irreducible (analogous to the requirement
for an induced group representation that the representation of the
little group $H$ on $V$ is irreducible). However, there is no
reason to expect that these conditions are sufficient to guarantee
irreducibility for a general category $\Q$, and it seems likely
that this issue has to be settled with a case-by-case study for
categories of physical importance.

\section{When $\Q$ is a Category of Sets}
\label{Sec:CategoryOfSets}
\subsection{Finite or countably infinite sets}
\label{SubSec:FiniteOrCountableSets} In practice, many of the
categories encountered in physics have objects that are {\em
sets\/} with internal structure, such as causal sets or
topological spaces. It is important, therefore, to see how the
general quantisation scheme outlined above can be explicitly
implemented in this case. The key questions to be addressed are:
\begin{enumerate}
    \item[i)] Is there a natural choice for the Hilbert spaces
    $\K(A)$, $A\in\Ob\Q$?
    \item[ii)] Given such a choice, can we find linear maps
    $\kappa(f):\K(B)\rightarrow\K(A)$, where
    $f:A\rightarrow B$, that satisfy the (presheaf) conditions in
    \eq{kcondPS}?
\end{enumerate}

To make life simpler, we start with the case in which $\Q$ is a
category of {\em finite\/} sets; for example, as in Figure
\ref{Fig:Cat4objects}.

A natural vector space associated with any set $A$  is the space
$F(A,\mathC)$ of all complex-valued functions on $A$. If
$A=\{a_1,a_2,\ldots,a_n\}$ is a {\em finite\/} set, then each
function $f:A\rightarrow\mathC$ leads to an $n$-tuple of complex
numbers, namely $(f(a_1),f(a_2),\ldots,f(a_n))$. Conversely, to
each $n$-tuple of complex numbers $(c_1,c_2,\ldots,c_n)$ there is
associated a function $f:A\rightarrow\mathC$, defined by
$f(a_i):=c_i$ for all $i=1,2,\ldots,n$. Thus we have an
isomorphism $F(A,\mathC)\simeq\mathC^{|A|}$, where $|A|$ denotes
the number of elements in the finite set $A$. Furthermore, since
$\mathC^n$ is a Hilbert space for any $n<\infty$, we can  make
$F(A,\mathC)$ into a Hilbert space by defining the inner product:
\begin{equation}
    \langle f,g\rangle:=\sum_{a\in A}f^*(a)g(a).
                \label{Def:IPFiniteA}
\end{equation}

If $A$ is countably infinite,  the inner product in
\eq{Def:IPFiniteA} can still be used provided we restrict our
attention to square-summable functions $f$; \ie, functions for
which   the sum
\begin{equation}
    \parallel f\parallel :=
            \left\{\sum_{a\in A}|f(a)|^2\right\}^{1/2}
\end{equation}
converges. We shall denote this Hilbert space by $\ell^2(A)$.

Thus if $A$ is finite or countably infinite, there is a natural
choice for the Hilbert space $\K(A)$: namely, $\mathC^{|A|}$ or
$\ell^2(A)$ respectively. The next thing to consider is if
$f:A\rightarrow B$ is a function between sets $A$ and $B$, is
there some natural induced map $\kappa(f):\K(B)\rightarrow\K(A)$?

In the case of a category of finite sets, there is indeed such a
map. For if $v:B\rightarrow \mathC$ belongs to
$\K(B)\simeq\mathC^{|B|}$, then, using the diagram
$A\mapright{f}B\mapright{v}\mathC$, we see that the natural
`pull-back' of $v:B\rightarrow\mathC$ is the complex-valued
function $v\circ f$ on $A$. Thus we try defining
$\kappa(f):\mathC^{|B|}\rightarrow \mathC^{|A|}$ by
\begin{equation}
        (\kappa(f)v)(a):=v(f(a))    \label{Def:kappa(f)}
\end{equation}
for all $a\in A$. Then, if $A\mapright{f}B\mapright{g}C$, and if
$v:C\rightarrow\mathC$, we have
\begin{equation}
        (\kappa(g\circ f)v)(a)=v(g\circ f(a))=v(g(f(a))
\end{equation}
whereas
\begin{equation}
    (\kappa(f)\kappa(g)v)(a)=(\kappa(g)v)(f(a))=v(g(f(a))),
\end{equation}
and hence $\kappa(f)\kappa(g)=\kappa(g\circ f)$, as required.

One could try to do the same thing in the case where there are a
countably infinite number of objects, with the Hilbert spaces
$\K(A)$, $A\in\Ob\Q$, chosen to be $\ell^2(A)$. However, a more
careful analysis of the situation is required now, since a
square-summable sequence (\ie, an element of $\ell^2(B)$) may not
be taken into another such sequence by the map $\kappa(f)$ defined
in \eq{Def:kappa(f)}.

\subsection{A very simple example}
\begin{figure}[h]
\begin{picture}(100,70)(-70,0)
\put(60,12){
\begin{picture}(80,100)(0,0)
    \put(0,0){$\bullet\ a $}\put(-2,-13){$A$}
\end{picture}
\begin{picture}(80,100)(0,0)
\put(0,0){$\bullet\ b_1$} \put(2.5,3){\vector(0,1){40}}
\put(0,42){$\bullet\ b_2$}\put(-2,-13){$B$}
\end{picture}
}
\end{picture}
\caption[]{A collection of two causal sets}\label{Fig:Cat2objects}
\end{figure}
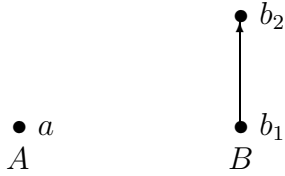

We will now illustrate the general scheme with the aid of a very
simple example. This is a category with just two objects: the
causal sets $A:=\{a\}$, and $B:=\{b_1,b_2\}$ with $b_1\leq b_2$.
This is shown in Figure \ref{Fig:Cat2objects}.

The arrows are order-preserving maps, and there are two such,
$f_1$, $f_2$, from $A$ to $B$ defined by
\begin{equation}
            f_1(a)=b_1
\end{equation}
and
\begin{equation}
            f_2(a):=b_2
\end{equation}
respectively, and one arrow $g$ from $B$ to $A$ defined by
\begin{equation}
            g(b_1):=a,\ g(b_2):=a.
\end{equation}

A non-trivial arrow $r:B\rightarrow B$ is
\begin{equation}
    r(b_1):=b_1,\ r(b_2):=b_1
\end{equation}
as is  $s:B\rightarrow B$ defined by
\begin{equation}
    s(b_1):=b_2,\ s(b_2):=b_2.
\end{equation}

If we forget the causal structure on $B$,  an additional arrow
$p:B\rightarrow B$ is
\begin{equation}
p(b_1):=b_2,\ p(b_2):=b_1           \label{Def:mapp}
\end{equation}
which is the non-trivial element of the permutation group
$\mathZ_2$ of the set $B$. It is not, however, an arrow in the
category $\Q$ since it reverses the ordering of the elements
$b_1,b_2\in B$.

In summary, we have the following sets of arrows:
\begin{eqnarray}
        \Hom{A,B}&=&\{f_1,f_2\}     \\
        \Hom{B,A}&=&\{g\}           \\
        \Hom{A,A}&=&\{\id_A\}       \\
        \Hom{B,B}&=&\{\id_B,r,s\}
\end{eqnarray}
to which should be added the map $p:B\rightarrow B$ in
\eq{Def:mapp} if we forget the causal structure on $B$.

The discussion in Section \ref{SubSec:FiniteOrCountableSets}
suggests that the appropriate choices for the Hilbert space fibres
are $\K[A]=\mathC^1\simeq\mathC$, and $\K[B]=\mathC^2$. Thus the
quantum state space of the system is
$\mathC\oplus\mathC^2\simeq\mathC^3$; a vector $\psi$ in this
space will be denoted by
$(\psi_A;\psi_{B_1},\psi_{B_2})\in\mathC^3$.

The expression \eq{Def:kappa(f)} for the multiplier, gives the
following representations \eq{Def:a(X)PS} of the non-trivial arrow
fields $X_f$, $f\in\Hom\Q$, where we denote $\hat a(X_f)$ by $\hat
a(f):=\hat a(X_f)$:
\begin{eqnarray}
\hat a(f_1):(\psi_A;\psi_{B_1},\psi_{B_2})&\mapsto&
            (\psi_{B_1};\psi_{B_1},\psi_{B_2})\label{af1} \\
\hat a(f_2):(\psi_A;\psi_{B_1},\psi_{B_2})&\mapsto&
            (\psi_{B_2};\psi_{B_1},\psi_{B_2})\\
\hat a(g):(\psi_A;\psi_{B_1},\psi_{B_2})&\mapsto&
            (\psi_A;\psi_A,\psi_A)\\
\hat a(r):(\psi_A;\psi_{B_1},\psi_{B_2})&\mapsto&
            (\psi_A;\psi_{B_1},\psi_{B_1})\\
\hat a(s):(\psi_A;\psi_{B_1},\psi_{B_2})&\mapsto&
            (\psi_A;\psi_{B_2},\psi_{B_2}).\label{as}
\end{eqnarray}
Note that arrows with the same domain and range are indeed
separated.

Of course, it is easy to write these operators as $3\times 3$
matrices. For example,
\begin{equation}
    \hat a(f_1):\pmatrix{a\cr b\cr c}\mapsto
    \pmatrix{b\cr b\cr c}=\pmatrix{0&1&0\cr
    0&1&0\cr 0&0&1}\pmatrix{a\cr b\cr c};
\end{equation}
so that
\begin{equation}
    \hat a(f_1)=
    \pmatrix{0&1&0\cr 0&1&0\cr 0&0&1}.
\end{equation}

The complete set of operators is
\begin{eqnarray}
    \hat a(f_1)&=& \pmatrix{0&1&0\cr 0&1&0\cr 0&0&1},\hspace{20pt}
        \hat a(f_2)=\pmatrix{0&0&1\cr 0&1&0\cr 0&0&1}\\[5pt]
    \hat a(g)&=& \pmatrix{1&0&0\cr 1&0&0\cr 1&0&0} \nonumber\\[5pt]
    \hat a(r)&=& \pmatrix{1&0&0\cr 0&1&0\cr 0&1&0},\hspace{20pt}
        \hat a(s)=\pmatrix{1&0&0\cr 0&0&1\cr 0&0&1}\nonumber\\[5pt]
    \hat \beta\ &=&
     \pmatrix{\beta_1&0&0\cr 0&\beta_2&0\cr 0&0&\beta_2} \nonumber
\end{eqnarray}
where $\beta_1:=\beta(A)$ and $\beta_2:=\beta(B)$.

 If we forget the causal structure on the set $B$, then there
is the additional arrow $p:B\rightarrow B$ defined in
\eq{Def:mapp}. The corresponding arrow field $X_p$ is represented
in the quantum theory by the matrix
\begin{equation}
\hat a(p)= \pmatrix{1&0&0\cr 0&0&1\cr 0&1&0}.
\end{equation}

\section{Conclusions}
We have seen how to construct a quantum scheme for a system whose
configuration space (or history analogue) is the set of object
$\Ob\Q$ in a small category $\Q$. A key ingredient is the monoid
$\AF\Q$ of arrow fields and its action on $\Ob\Q$. Multiplier
representations are needed to distinguish quantum theoretically
between arrows with the same range and domain. Each such
representation can be expressed in terms of a presheaf of Hilbert
spaces over $\Ob\Q$. For the example of a category of finite sets
we have shown how to construct an explicit example of such a
presheaf.

This scheme includes as a special case the situation in which the
configuration space is a manifold $Q$ on which a Lie group $G$
acts transitively, so that $Q\simeq G/H$ for some subgroup $H$ of
$G$. In this sense, the scheme can be viewed as a big extension of
standard quantisation methods to include types of system whose
configuration spaces (or history analogue) are far from being
points in a smooth manifold.

It is an exciting challenge to use these techniques for
constructing novel theories of quantised space or space-time.
However, it should be emphasised that what is described in the
present paper is only a `tool-kit' for constructing such theories:
it needs a creative leap to use these tools to construct a
physically realistic model of, for example, quantum causal sets.
The key step would be to choose a decoherence functional for the
quantum history theory. This decoherence functional would be
constructed from the operators described in this paper, but new
physical principles are needed to decide its precise form. This is
an important topic for future research.

\section*{Acknowledgments}

\noindent Support by the EPSRC in the form of the grant GR/R36572
is gratefully acknowledged.

\end{document}